\documentclass[apj]{emulateapj}

\input{epsf}

\begin{document}
\def\lax    {\ifmmode{_<\atop^{\sim}}\else{${_<\atop^{\sim}}$}\fi}
\def\gax    {\ifmmode{_>\atop^{\sim}}\else{${_>\atop^{\sim}}$}\fi}
\def\gtorder{\mathrel{\raise.3ex\hbox{$>$}\mkern-14mu
             \lower0.6ex\hbox{$\sim$}}}
\def\ltorder{\mathrel{\raise.3ex\hbox{$<$}\mkern-14mu
             \lower0.6ex\hbox{$\sim$}}}
 

 
\title{Chandra and XMM-Newton discovery of the transient X-ray pulsar in the nearby spiral galaxy NGC 2403}


\author{Sergey P. Trudolyubov\altaffilmark{1,2}, William C. Priedhorsky\altaffilmark{3}, 
and France A. C\'ordova\altaffilmark{1,4}}

\altaffiltext{1}{Institute of Geophysics and Planetary Physics, University of 
California, Riverside, CA 92521}

\altaffiltext{2}{Space Research Institute, Russian Academy of Sciences, 
Profsoyuznaya 84/32, Moscow, 117810 Russia}

\altaffiltext{3}{Los Alamos National Laboratory, Los Alamos, NM 87545}

\altaffiltext{4}{Department of Physics and Astronomy, University of California, Riverside, CA 92521}

\begin{abstract}
We report on the discovery and analysis of the transient X-ray pulsar CXOU J073709.1+653544 detected in the 2004 
August--October {\em Chandra} and {\em XMM-Newton} observations of the nearby spiral galaxy NGC 2403. The X-ray source 
exhibits X-ray pulsations with a period P$\sim$18 s and a nearly sinusoidal pulse shape and pulsed fraction 46-70\% 
during the first three observations. A detailed timing analysis reveals a rapid decrease of the pulsation 
period from 18.25 s on Aug. 9 to 17.93 s on Sep. 12 and possibly 17.56 s on Oct. 3, 2004. The X-ray spectra of 
CXOU J073709.1+653544 are hard and are well fitted with an absorbed simple power law of photon index 
$\Gamma \sim 0.9-1.2$ in the 0.3-7 keV energy band. The X-ray properties of the source and the absence of an optical/UV 
counterpart brighter than 20$^{\rm th}$ magnitude allow us to identify CXOU J073709.1+653544 as accreting X-ray pulsar 
in NGC 2403. The maximum unabsorbed luminosity of the source in the 0.3-7 keV range, ${\rm L}_{\rm X} \sim 2.6\times 10^{38}$ 
ergs s$^{-1}$ at 3.2 Mpc, is at least 260 times higher than its quiescent luminosity, and exceeds the isotropic Eddington 
limit for a 1.4${\rm M}_{\odot}$ compact object accreting hydrogen-rich material. The corresponding estimated luminosity 
in the 0.3-100 keV energy range could be as high as $\sim 1.2\times10^{39}$ ergs s$^{-1}$, assuming the typical pulsar 
energy spectrum with high-energy cut-off at 10-20 keV. The rate of decrease of the pulsation period of the source 
($\dot{P}\sim-10^{-7}$ s s$^{-1}$) is one of the fastest observed among accreting pulsars. The evolution of the pulsation 
period suggests that it is dominated by the intrinsic spin-up of the compact object, which is almost certainly a neutron star. 
The X-ray luminosity of CXOU J073709.1+653544 is high enough to account for the observed spin-up rate, assuming that the X-ray 
source is powered by disk accretion onto highly magnetized neutron star. Based on the transient behavior and overall X-ray 
properties of the source, we conclude that it could be an X-ray pulsar belonging to either a Be binary system or a low-mass 
system similar to the transient Galactic bursting pulsar GRO J1744-28.     
\end{abstract}


\section{INTRODUCTION}
The observations of accretion-powered X-ray pulsars provide a unique opportunity to study the dynamical properties of 
accreting matter and its interaction with the magnetosphere of a neutron star, and the transfer of the angular momentum 
in a binary system (White, Swank \& Holt 1983; Nagase 1989; Bildsten et al. 1997). Most of the known accreting X-ray 
pulsars are found in the high-mass binary systems involving a neutron star and OB donor star (either supergiant or Be 
type) and are associated with younger stellar populations \cite{Corbet86,CC06}. These two types correspond to different 
modes of accretion with supergiant systems accreting from a radially outflowing stellar wind, and the Be binaries accreting 
from a circumstellar disk. Therefore, the accreting pulsars with supergiant companions are observed as persistent X-ray 
sources, while the Be systems are recurrent transients and highly variable sources and reach much higher X-ray luminosities 
up to $10^{38}\sim10^{39}$ ergs s$^{-1}$. A small number of X-ray pulsars is associated with low-mass binary systems with 
mass donors ranging from a main-sequence stars to degenerate carbon-oxygen dwarfs and red giants and can reach persistent 
luminosities of $\gtrsim 10^{38}$ ergs s$^{-1}$ (Nagase 1989; Bildsten et al. 1997).   
            
Until recently, the study of X-ray pulsars was limited to our Galaxy and the nearby Magellanic Clouds due to the limited 
sensitivity and spatial resolution of previous X-ray missions. With the advent of the {\em Chandra} and {\em XMM-Newton} 
X-ray observatories, it has become possible to study the spectral and timing properties of individual X-ray sources 
associated with more distant galaxies, including bright X-ray pulsators \cite{FW06,F06}. For example, recent {\em XMM-Newton} 
observations of M31 revealed a 865 s pulsating supersoft X-ray source and a 197 s accreting X-ray pulsar candidate with 
luminosities of $\sim 10^{37}$ ergs s$^{-1}$ and pulsed fractions of $\sim 40\%$ \cite{O01,T05}. These results demonstrate 
that bright pulsating X-ray sources ($L_{X}\gtrsim 10^{38}$ ergs s$^{-1}$) can be detected with {\em XMM-Newton} and 
{\em Chandra} in the galaxies beyond the Local Group up to the distances of a few Mpc. 

The spiral Scd galaxy NGC 2403 at 3.2 Mpc (Freedman \& Madore 1988), provides a good opportunity to study X-ray source 
populations in a normal galaxy. NGC 2403 was a target of previous X-ray observations with the {\em Einstein} (Fabbiano \& 
Trinchieri 1987; Fabbiano et al. 1992), {\em ROSAT} (Roberts \& Warwick 2000; Liu \& Bregman 2005) and {\em ASCA} 
observatories (Kotoku et al. 2000), which detected five discrete sources with one bright source classified as an 
ultraluminous X-ray source (ULX) with luminosity above $10^{39}$ ergs s$^{-1}$. Recent {\em Chandra} observations covering 
the inner $\sim 25\%$ of the $D_{25}$ area of NGC 2403 have revealed 41 discrete sources with apparent luminosities down to 
$\sim 10^{36}$ ergs s$^{-1}$ and allowed to study unresolved soft X-ray emission from the galactic disk \cite{SP03,F02}. 
In addition, the survey of luminous X-ray source populations with {\em XMM-Newton} allowed to study spectral properties of 
four brightest sources detected in NGC 2304 (Winter et al. 2006)      
 
In this paper, we report on the discovery of the transient pulsating X-ray source CXOU J073709.1+653544 in NGC 2403 
using publicly available August-October 2004 observations made with {\em Chandra} and {\em XMM-Newton} as a part of the 
follow-up program studying recent supernova SN 2004dj \cite{PL04}.     

\section{OBSERVATIONS AND DATA REDUCTION}
In our analysis we used the data of four 2004 August-December {\em Chandra}/ACIS-S observations of SN 2004dj region and 
2001 April 17 observation of NGC 2403 (Table \ref{obslog}). The data of {\em Chandra} observations was processed using the 
CIAO v3.3\footnote{http://asc.harvard.edu/ciao/} threads. We performed standard screening of the {\em Chandra} data to 
exclude time intervals with high background levels, and found no flares exceeding 20\% of the average background level. For 
each observation, we generated X-ray images in the 0.3-7 keV energy band, and used CIAO wavelet detection routine 
{\em wavdetect} to detect point sources. Only data in the 0.3-7 keV energy range was used in the spectral analysis of ACIS-S 
observations. The standard CIAO {\em axbary} tool was used to produce a barycenter-corrected source event lists used in the 
timing analysis. During three 2004 August-October observations the source was positioned on the back-illuminated chip S3, 
offset by $1.3\arcmin\sim1.4\arcmin$ from the aim point. To produce source filtered event lists and spectra, we extracted counts 
within the inclined elliptical region with semi-axes of $2\arcsec$ and $2.5\arcsec$ centered at the position of CXOU J073709.1+653544, 
large enough to include more than 90\% of the source energy flux. Background counts were extracted from the adjacent source-free 
regions with subsequent normalization by ratio of detector areas. 

We also use the data of 2003 April 30, 2003 September 11 and 2004 September 12 {\em XMM-Newton} observations of NGC 2403 
(Table \ref{obslog}) with three European Photon Imaging Camera (EPIC) instruments (MOS1, MOS2 and pn)\cite{Turner01,Strueder01}, 
and the Optical Monitor (OM) telescope (Mason et al. 2001). We reduced {\em XMM} data using {\em XMM-Newton} Science Analysis 
System (SAS v 6.5.0)\footnote{See http://xmm.vilspa.esa.es/user}. For 2004 September 12 observation, we filtered the original 
EPIC event files based on the good time intervals produced from the light curves extracted from the whole detector in the 0.2-12 
keV energy band by applying an upper count rate threshold of 6 and 2 counts s$^{-1}$ for the EPIC pn and MOS cameras. The last 
$\sim 20$ ks of the EPIC observation is affected by high background, and has been excluded from analysis. The standard SAS tool 
{\em barycen} was used to perform barycentric correction on the original EPIC event files used for timing analysis. A significant 
part of 2003 April 30 and September 11 {\em XMM-Newton} observations was affected by background flares. After screening for high 
background only $4.7$ ks (2003 Apr. 30 observation) and $\sim 9$ ks (2003 Sep. 11 observation) of EPIC-MOS exposure and 5 ks of 
EPIC-pn exposure (2003 Sep. 11 observation) was left for the analysis (Table \ref{obslog}).     

We generated EPIC-pn and MOS images of NGC 2403 field in the 0.3-7.0 keV energy band, and used the SAS standard maximum 
likelihood (ML) source detection script {\em edetect\_chain} to detect point sources. We used bright X-ray sources with 
known counterparts from USNO-B catalog \cite{Monet03} and {\em Chandra} source lists to correct EPIC image astrometry. 
The astrometric correction was also applied to the OM images, using cross-correlation with USNO-B catalog. After 
correction, we estimate residual systematic error in the source positions to be of the order $0.5 - 1\arcsec$ for both 
EPIC and OM. 

To generate EPIC-MOS source lightcurves and spectra during the 2004 September 12 {\em XMM} observation, we used a circular region 
of $18\arcsec$ radius centered at the position of CXOU J073709.1+653544. Due to the source proximity to the edge of EPIC-pn CCD, 
the source counts were extracted from the elliptical region with semi-axes of $18\arcsec$ and $13\arcsec$ and position angle of 20 
degrees, including $\sim 70\%$ of the source energy flux. The adjacent source-free regions were used to extract background spectra and 
lightcurves. The source and background spectra were then renormalized by ratio of the detector areas. For spectral analysis, we used 
data in the $0.3 - 10$ keV energy band. To synchronize source and background lightcurves from individual EPIC detectors, we used the 
identical time filtering criteria based on Mission Relative Time (MRT), following the procedure described in Barnard et al. (2007). 
The background lightcurves were not subtracted from the source lightcurves, but were used later to estimate the background contribution 
in the calculation of the source pulsed fractions.

To estimate upper limits on the quiescent source luminosities, the ACIS and EPIC count rates were converted into energy fluxes 
in the 0.3-7 keV energy range using Web PIMMS\footnote{http://heasarc.gsfc.nasa.gov/Tools/w3pimms.html}, assuming an absorbed 
power law spectral shape with photon index $\Gamma = 1.0$ and Galactic foreground absorbing column N$_{\rm H}$=$4\times10^{20}$ 
cm$^{-2}$ \cite{DL90}. 

The energy spectra were grouped to contain a minimum of 20 counts per spectral bin in order to allow $\chi^{2}$ statistics, 
and fit to analytic models using the XSPEC v.12\footnote{http://heasarc.gsfc.nasa.gov/docs/xanadu/xspec/index.html} fitting 
package \cite{arnaud96}. EPIC-pn, MOS1 and MOS2 data were fitted simultaneously, but with normalizations varying independently. 
For timing analysis we used standard XANADU/XRONOS v.5\footnote{http://heasarc.gsfc.nasa.gov/docs/xanadu/xronos/xronos.html} 
tasks.

In the following analysis we assume a distance of 3.2 Mpc for NGC 2403 (Freedman \& Madore 1988). All parameter errors quoted 
are 68\% ($1\sigma$) confidence limits. 

\section{RESULTS}
\subsection{Source Position}
\begin{figure}
\includegraphics[clip=true,scale=0.5]{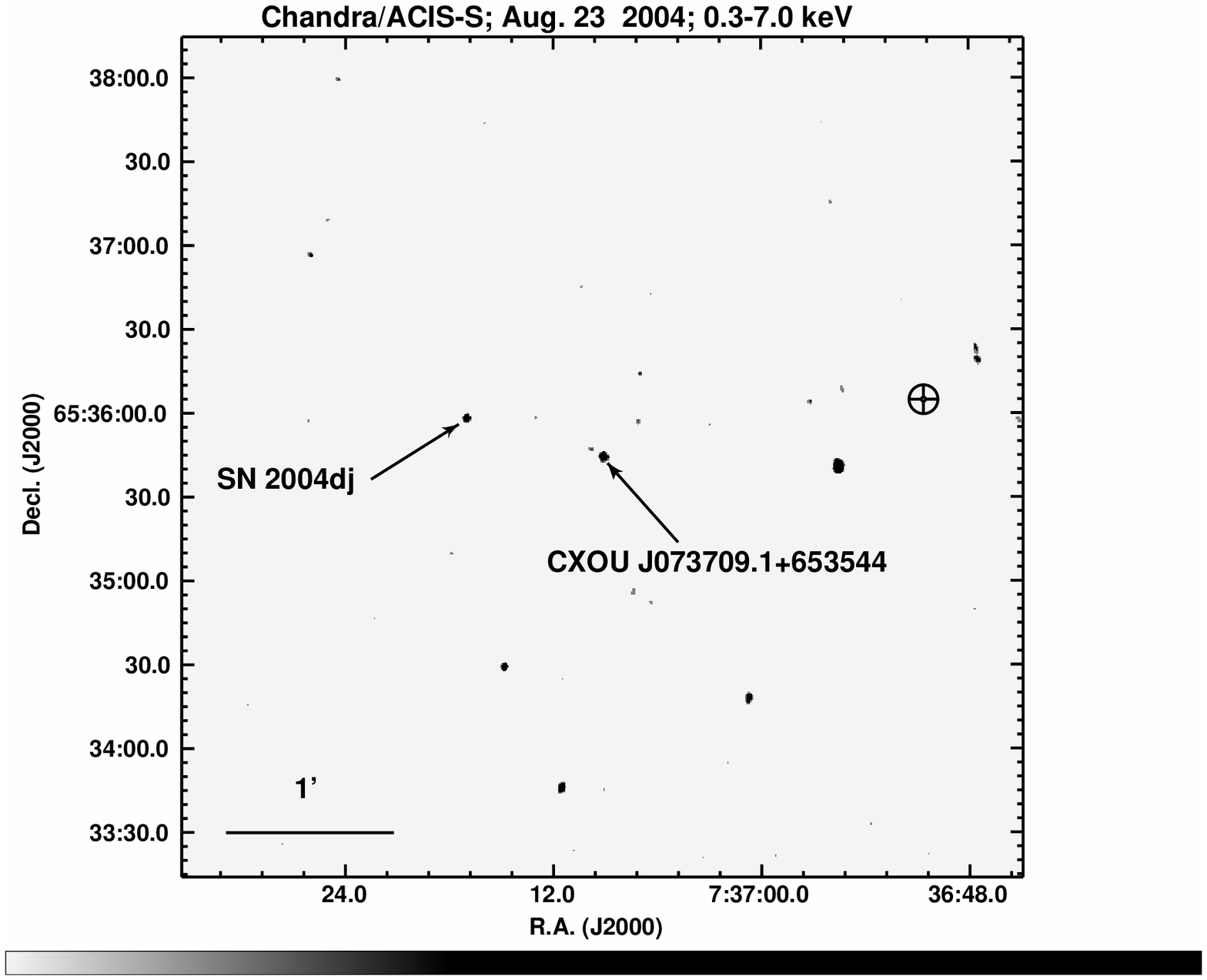}\includegraphics[clip=true,scale=0.5]{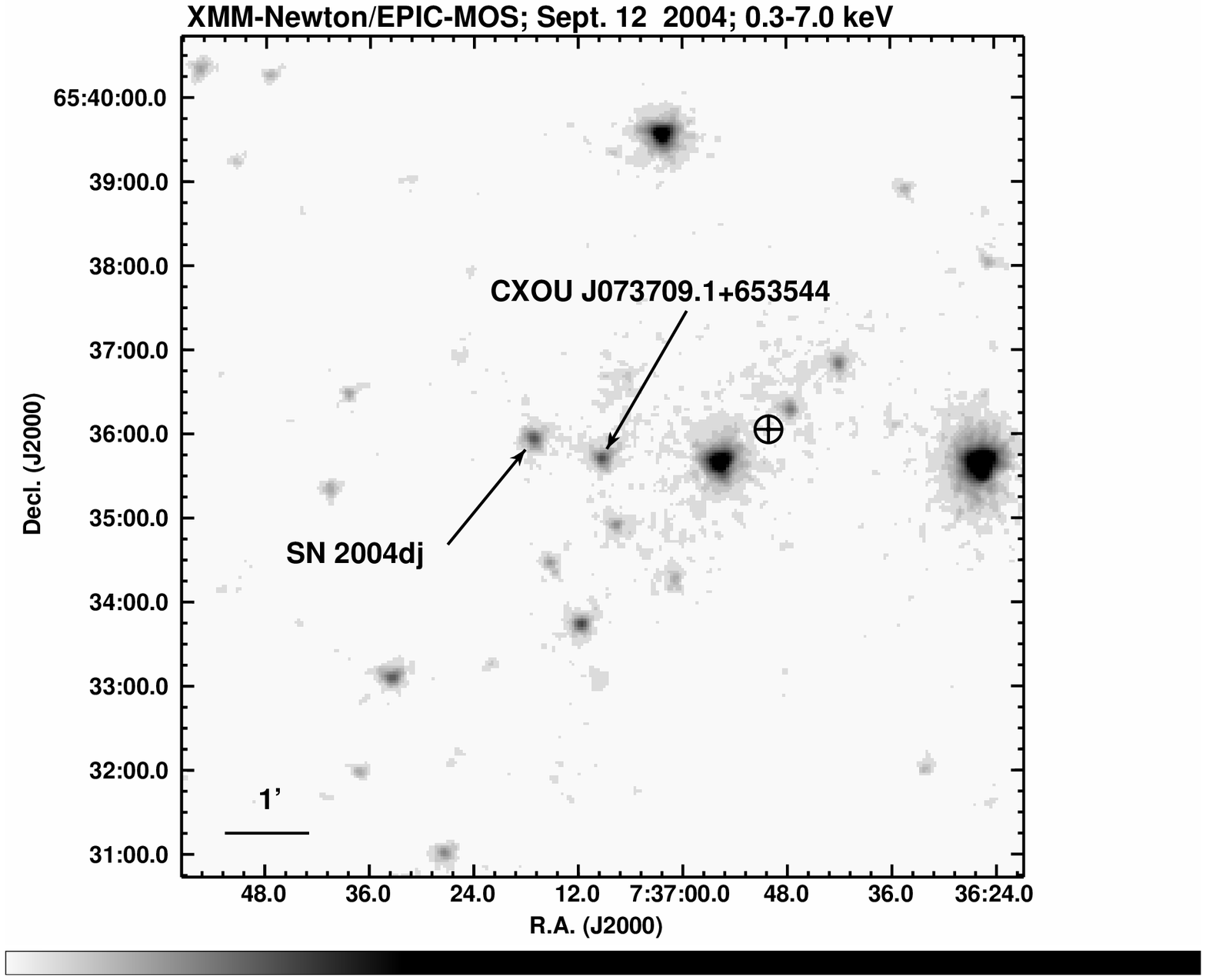}\\
\includegraphics[clip=true,scale=0.5]{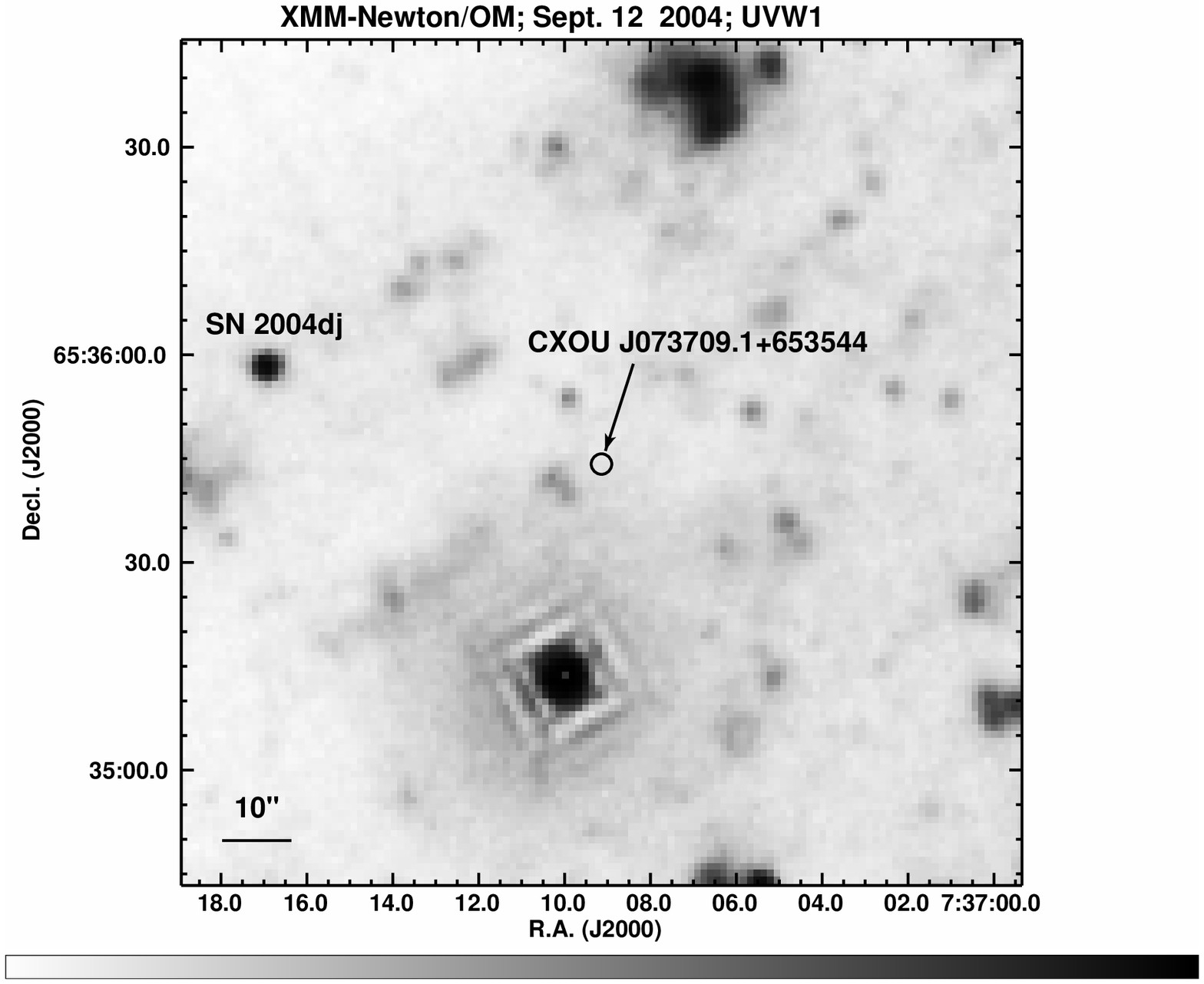}\includegraphics[clip=true,scale=0.5]{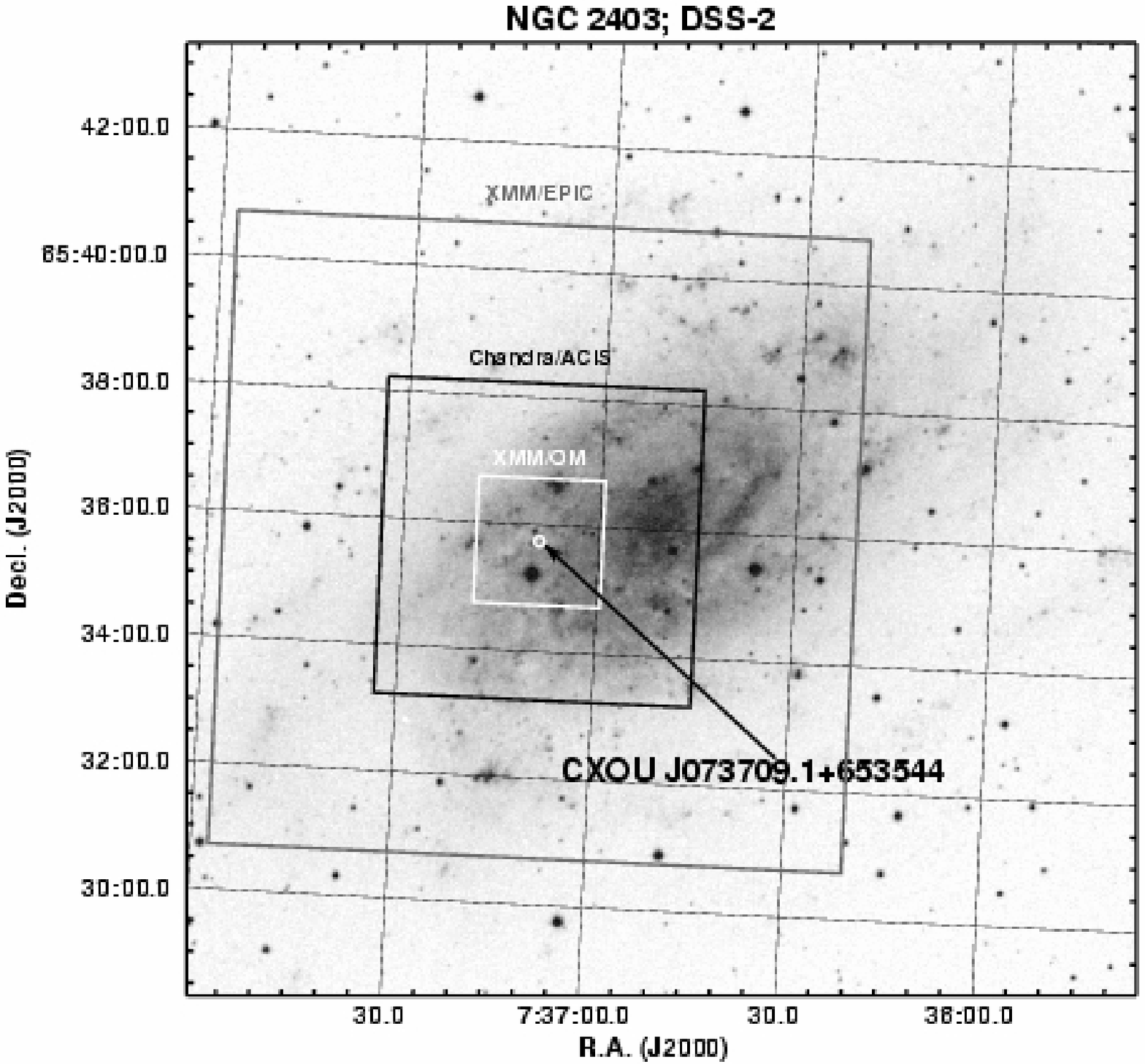}\\ 
\caption{{\em Upper left:} {\em Chandra}/ACIS-S image covering a 5$\arcmin\times5\arcmin$ region of NGC 2403 centered on the CXOU J073709.1+653544, taken on Aug. 23, 2004. {\em Upper right:} Combined 0.3-7 keV {\em XMM}/EPIC-MOS image covering a 10$\arcmin\times10\arcmin$ region of NGC 2403 centered on the CXOU J073709.1+653544 position, taken on Sept. 12, 2004. In both upper panels the positions of a new pulsar CXOU J073709.1+653544 and supernova SN 2004dj are marked with arrows, and the position of NGC 2403 nucleus is shown with a cross. {\em Lower left:} XMM-Newton/OM UVW1 band image of NGC 2403 field taken on Sept. 12, 2004. The image is a 2$\arcmin\times2\arcmin$ square centered on the CXOU J073709.1+653544 position. The localization of CXOU J073709.1+653544 is shown with black circle of $1.5\arcsec$ radius ($3\sigma$). {\em Lower right:} Optical B-band image of NGC 2403 from the Second Generation Digitized Sky Survey. The position of CXOU J073709.1+653544 is shown with white circle. The regions covered by {\em Chandra}/ACIS, {\em XMM}/EPIC and {\em XMM}/OM images are shown with black, gray and white squares. \label{image_general}}
\end{figure}

We discovered a new X-ray source CXOU J073709.1+653544 in the data of the 2004 August 9 and detected it in 
the subsequent August 23 and October 3 {\em Chandra} observations of the NGC 2403 field (Table \ref{obslog}). 
In addition, the analysis of archival 2004 September 12 {\em XMM-Newton} observation revealed the presence of 
a bright X-ray source at the position consistent with CXO J073709.13+653544. The source was not detected in 
the 2004 December 22 {\em Chandra} observation (observation ID 4630). Combining the data of three {\em Chandra} 
observations, we measure the position of CXOU J073709.1+653544 to be $\alpha = 07^{h} 37^{m} 09.139^{s}, 
\delta = +65\arcdeg 35\arcmin 44.23\arcsec$ (J2000 equinox) with an uncertainty of $\sim0.5\arcsec$ (Fig. 
\ref{image_general}). The projected galactocentric distance of CXOU J073709.1+653544 is $\sim 2\arcmin$ or 
$\sim 1.86$ kpc at 3.2 Mpc. 

The search for the optical counterparts using the images from the Digitized Sky Survey did not yield any 
stellar-like object brighter than $m_{\rm B} \sim 20$ within the error circle of CXOU J073709.1+653544. We 
used the data of 2003 April 30 and 2004 Sept. 12 {\em XMM-Newton}/OM observations to search for UV counterparts to 
the source prior and during its X-ray outburst (Fig. \ref{image_general}). We did not detect any stellar counterparts 
to CXOU J073709.1+653544 in the OM images down to the limit of $\sim 20^{m}$ in the U and OM UVW1 (291 nm) 
bands.

\subsection{Long-Term Flux Evolution}
\begin{figure}
\epsscale{1.0}
\plotone{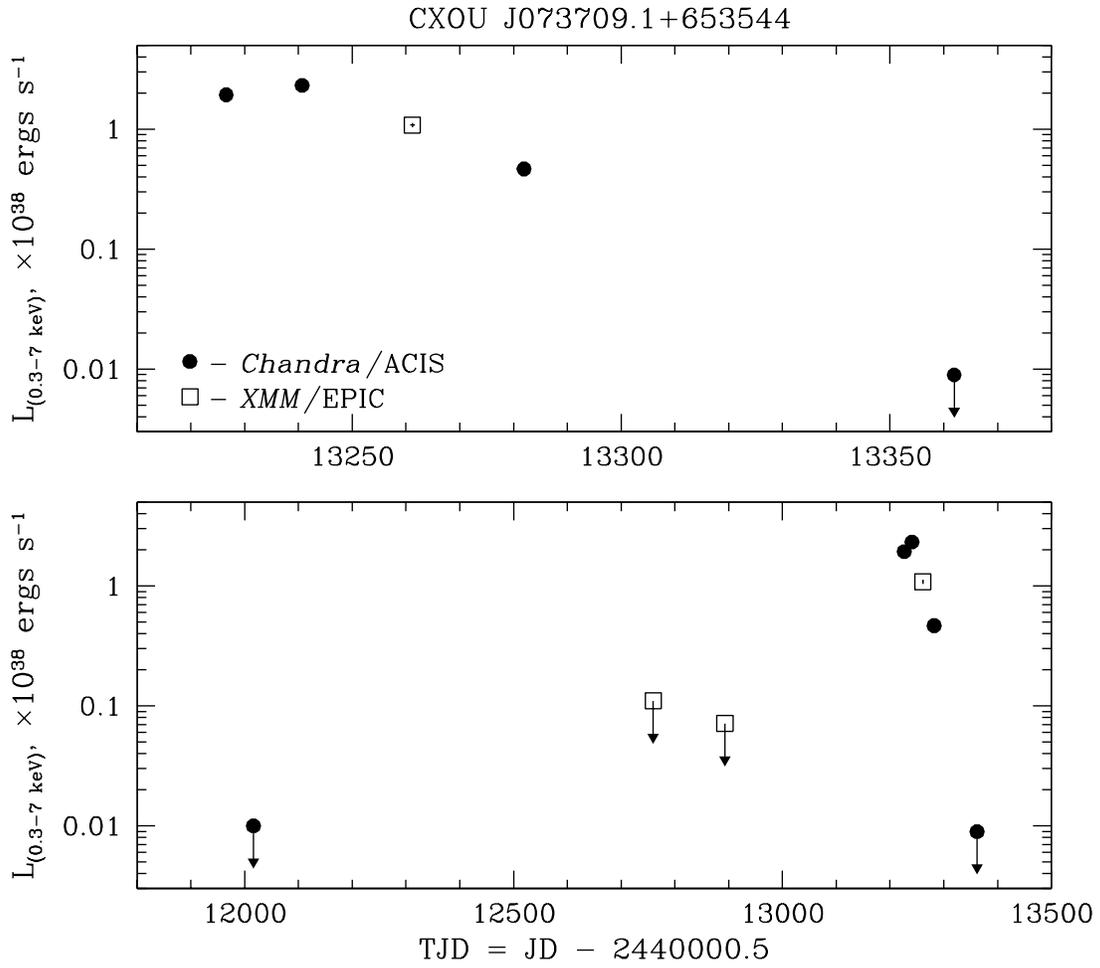}
\caption{(Upper panel:) The X-ray light curve of CXOU J073709.1+653544 in the 0.3-7 keV energy band based on the results of 2004 {\em Chandra} and {\em XMM-Newton} observations. (Lower panel:) long-term light curve of CXOU J073709.1+653544 based on the results of 2001-2005 {\em Chandra} and {\em XMM-Newton} observations. The luminosities were calculated assuming a source distance of 3.2 Mpc. \label{lc_long_term}}
\end{figure}

The evolution of X-ray luminosity of CXOU J073709.1+653544 in the 0.3-7 keV energy band based on the data 
of 2004 {\em Chandra} and {\em XMM-Newton} observations is shown in Fig. \ref{lc_long_term}. 
The first two August 2004 {\em Chandra} observations show the overall increase of the source X-ray luminosity in 
the 0.3-7 keV energy band from $\sim 1.9\times10^{38}$ to $\sim 2.3\times10^{38}$ ergs s$^{-1}$. The subsequent 
{\em XMM-Newton} and {\em Chandra} observations revealed the gradual decrease of the source luminosity to 
$\sim 10^{38}$ ergs s$^{-1}$ on Sept. 12, and $\sim 5\times10^{37}$ ergs s$^{-1}$ on Oct. 3, 2004. The source 
was not detected during the 2004 December 22 {\em Chandra} observation with an upper limit on its 0.3-7 keV luminosity 
of $\lesssim 10^{36}$ ergs s$^{-1}$ ($2\sigma$), $>230$ times lower than maximum measured outburst luminosity. It should 
be noted that due to the sparse sampling the lightcurve of the source observed during 2004 August-October could be a 
result of either single long ($\gtrsim 60$ days) outburst or the superposition of a number of shorter outbursts poorly 
sampled in time. The analysis of archival 2001 April 17 {\em Chandra} observation (observation ID 2014) \cite{SP03} of 
NGC 2403 taken prior to the 2004 outburst, allow us to estimate the average upper limit ($2\sigma$) on the source quiescent 
luminosity to be $\sim 10^{36}$ ergs s$^{-1}$ in the $0.3 - 7$ keV energy band. The source was not detected in the two 
2003 April 30 and September 11 {\em XMM-Newton} observations (Table \ref{obslog}), but with a much less stringent upper 
limit on its 0.3-7 keV luminosity, $\lesssim 10^{37}$ ergs s$^{-1}$ (Fig. \ref{lc_long_term}).  

\subsection{X-ray Pulsations}
\begin{figure}
\epsscale{0.9}
\plotone{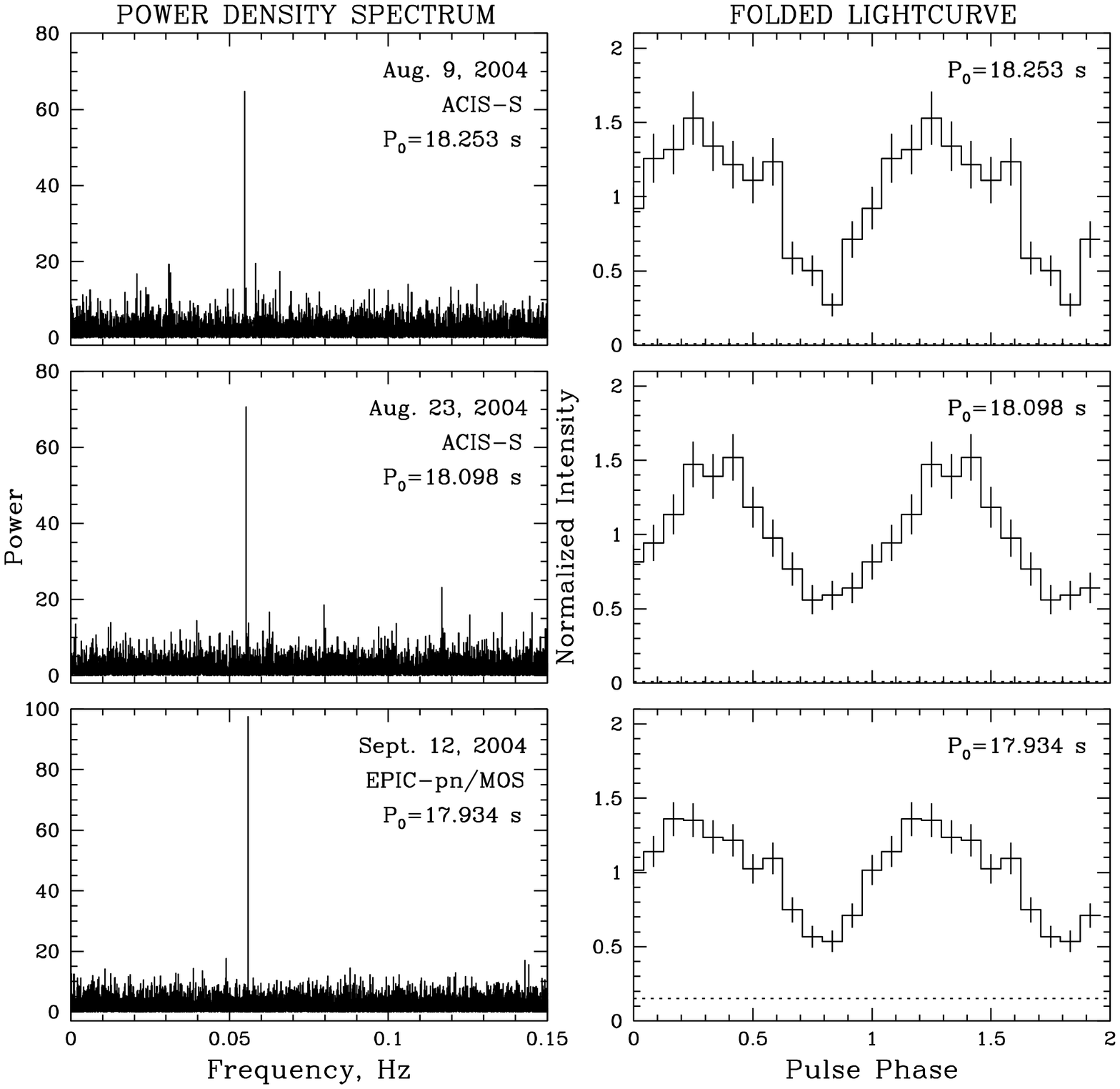}
\plotone{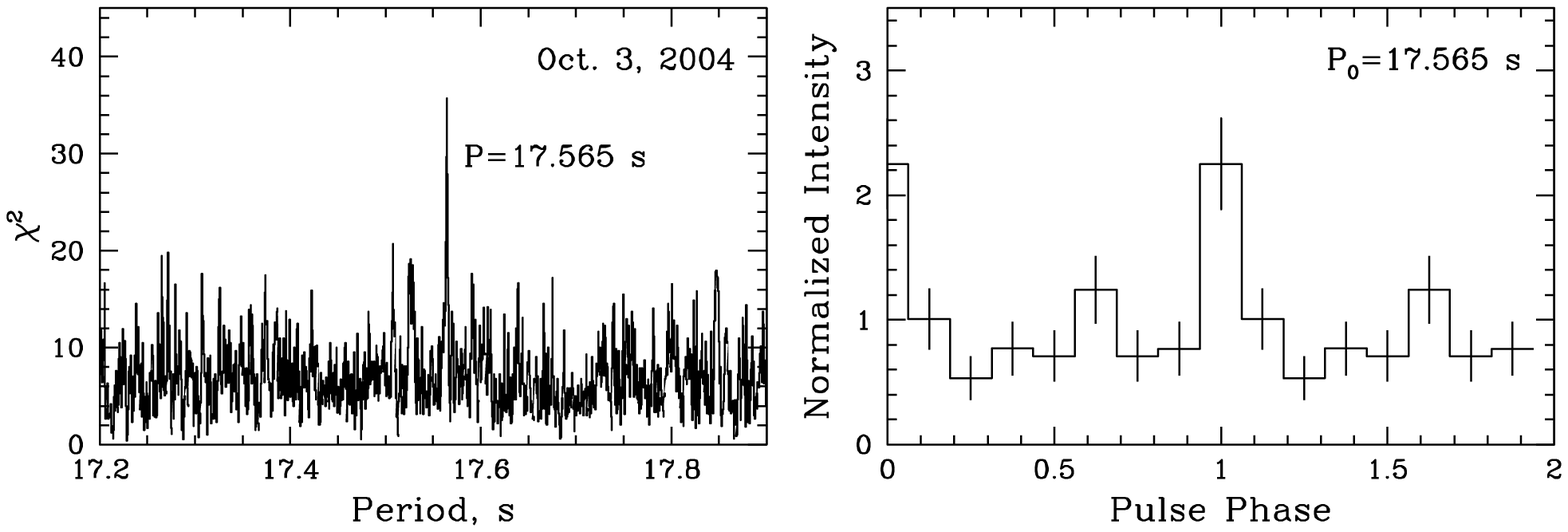}
\caption{(Left upper three panels) Power spectra of CXOU J073709.1+653544 obtained using the data of 2004 Aug. 9 
(upper panel) and Aug. 24 (middle panel) {\em Chandra}/ACIS-S observations and 2004 Sept. 12 {\em XMM-Newton}/EPIC 
observations (lower panel) in the 0.3-7 keV energy band. (Right upper three panels) Corresponding pulse profiles folded 
with most likely pulsation periods. The background levels are represented by the dotted lines. The ephemeris is defined 
arbitrarily such as the dip in the pulse profile falls at phase 0.8. (Lower panels) The results of epoch-folding analysis 
of the data of 2004 Oct. 3 {\em Chandra}/ACIS observation of the source. \label{pds_efold}}
\end{figure}

We performed a timing analysis of the CXOU J073709.1+653544 using the data from {\em Chandra}/ACIS-S and 
{\em XMM-Newton}/EPIC detectors in the 0.3-7 keV energy band. After a barycentric correction of the photon arrival 
times in the original event lists, we performed a Fast Fourier Transform (FFT) analysis using standard XRONOS task 
{\em powspec}, in order to search for coherent periodicities. For the analysis of {\em XMM-Newton} data, we used 
combined synchronized EPIC-pn and MOS lightcurves with 2.6 s time bins to improve sensitivity. We found strong peaks 
in the Fourier spectra of data from the 2004 Aug. 9 and 23 {\em Chandra} observations and 2004 Sept. 12 {\em XMM} 
observation at the frequencies of (5.48-5.58)$\times10^{-2}$ Hz (Fig. \ref{pds_efold}, {\em left panels}). The strengths 
of the peaks in the individual Fourier spectra (Fig. \ref{pds_efold}) correspond to the period detection confidence of 
$\sim 10^{-10}$, $\sim 10^{-11}$ and $\sim 10^{-17}$ for the Aug. 9, 23 {\em Chandra} and Sept. 12 {\em XMM-Newton} 
observations (disregarding simultaneous detection in all three observations). 

To estimate the pulsation periods more precisely, we used an epoch folding technique, assuming no period change during 
individual observations. The most likely values of the pulsation period (Table \ref{timing_spec_par}) were obtained 
fitting the peaks in the $\chi^{2}$ versus trial period distribution with a Gaussian. The period errors in Table 
\ref{timing_spec_par} were computed following the procedure described in Leahy (1987). Then the source lightcurves 
were folded using the periods determined from epoch folding analysis. The resulting folded lightcurves of CXOU J073709.1+653544 
in the 0.3-7 keV energy band during first three observations are shown in Fig. \ref{pds_efold} ({\em right panels}). The source 
demonstrates quasi-sinusoidal pulse profiles in the 0.3-7 keV energy band during the first three observations (Fig. \ref{pds_efold}). 
The pulsed fraction, defined as (I$_{\rm max}$-I$_{\rm min}$)/(I$_{\rm max}$+I$_{\rm min}$), where I$_{\rm max}$ and I$_{\rm min}$ 
represent source intensities at the maximum and minimum of the pulse profile excluding background photons, is significantly higher 
during the 2004 Aug. 9 {\em Chandra} observation (70$\pm$4\%) when compared to the observations \#5 (46$\pm$4\%) and \#6 
(56$\pm$3\%) (Table \ref{timing_spec_par}).  

Although the analysis of the Oct. 3 {\em Chandra} observation (obs. \#7) did not show the presence of a strong single 
peak in the power density spectrum, we still performed a search for the pulsations in the data of this observation 
using the epoch folding technique. The epoch folding analysis of the source lightcurve have led to the marginal detection 
of the pulsation with a period of 17.565 s (Fig. \ref{pds_efold}, {\em left panel}). The folded lightcurve of the 
source in the 0.3-7 keV energy band with a significance of $\sim 3.7\sigma$ determined using $\chi^{2}$ for a nonvarying, 
constant model, is shown in the lower right panel of Fig. \ref{pds_efold}. The pulse profile possibly shows a non-sinusoidal 
structure, which probably explains why we do not see strong first harmonic in its Fourier spectrum. 

\begin{figure}
\epsscale{0.30}\plotone{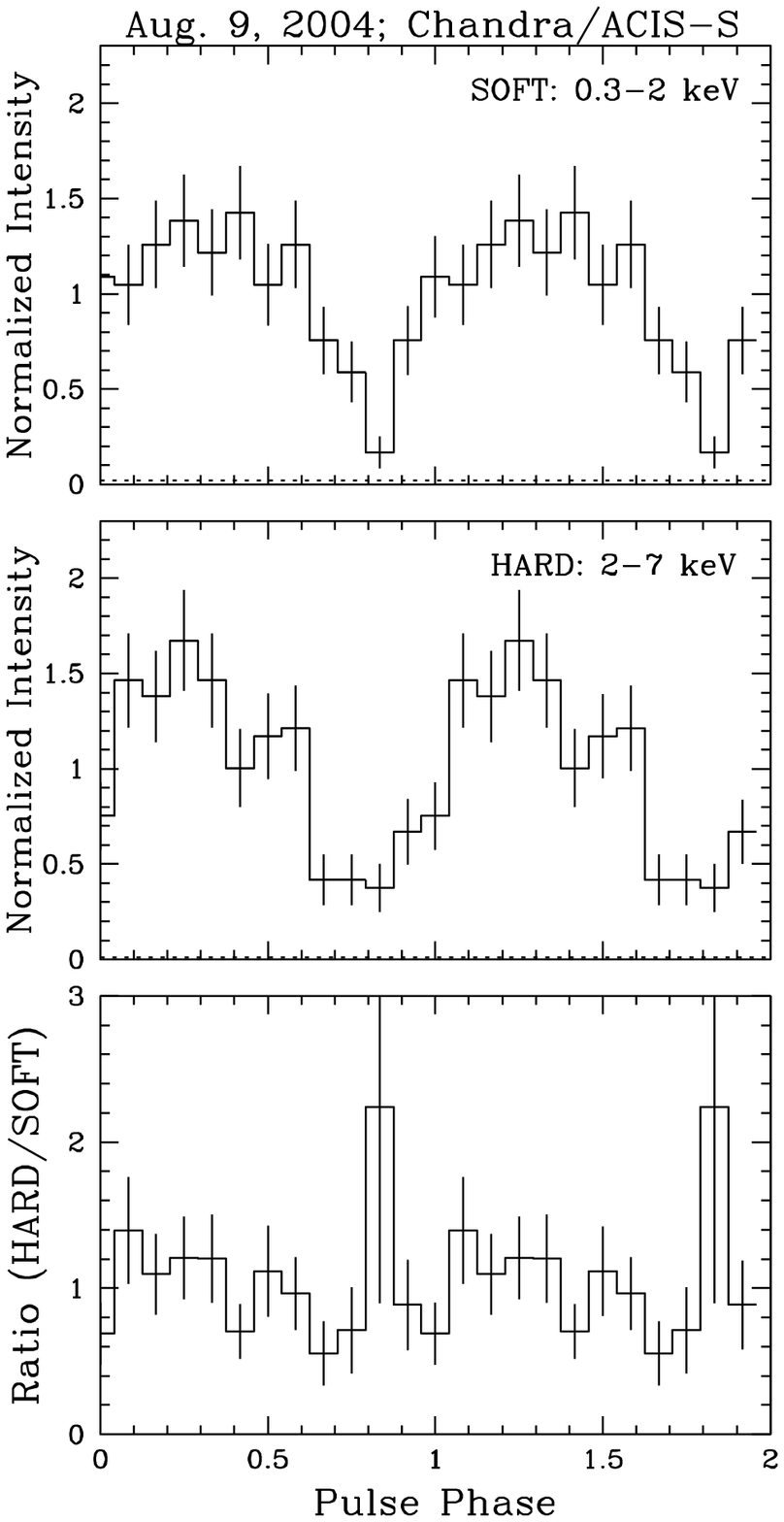}\epsscale{0.30}\plotone{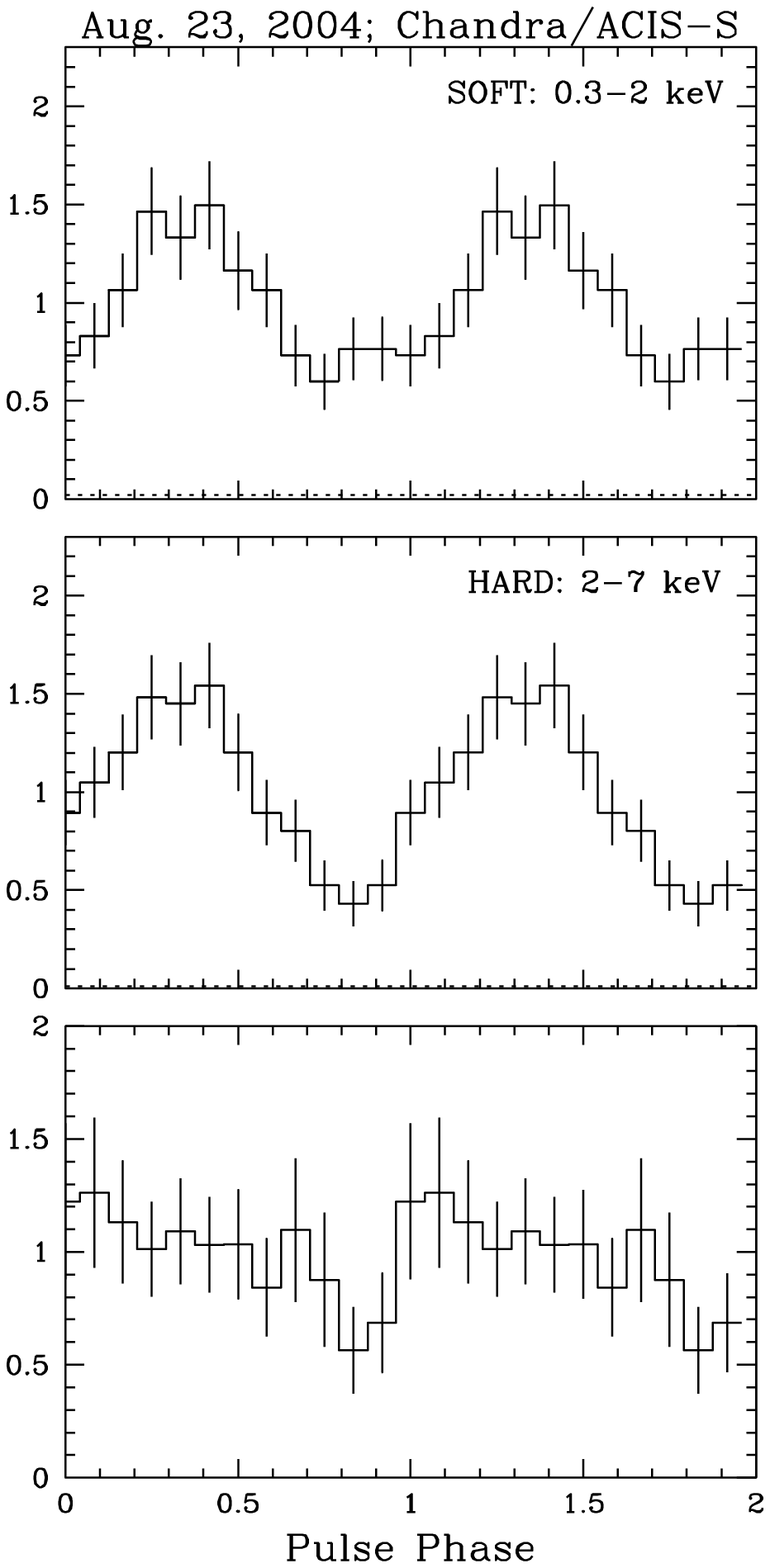}\epsscale{0.30}\plotone{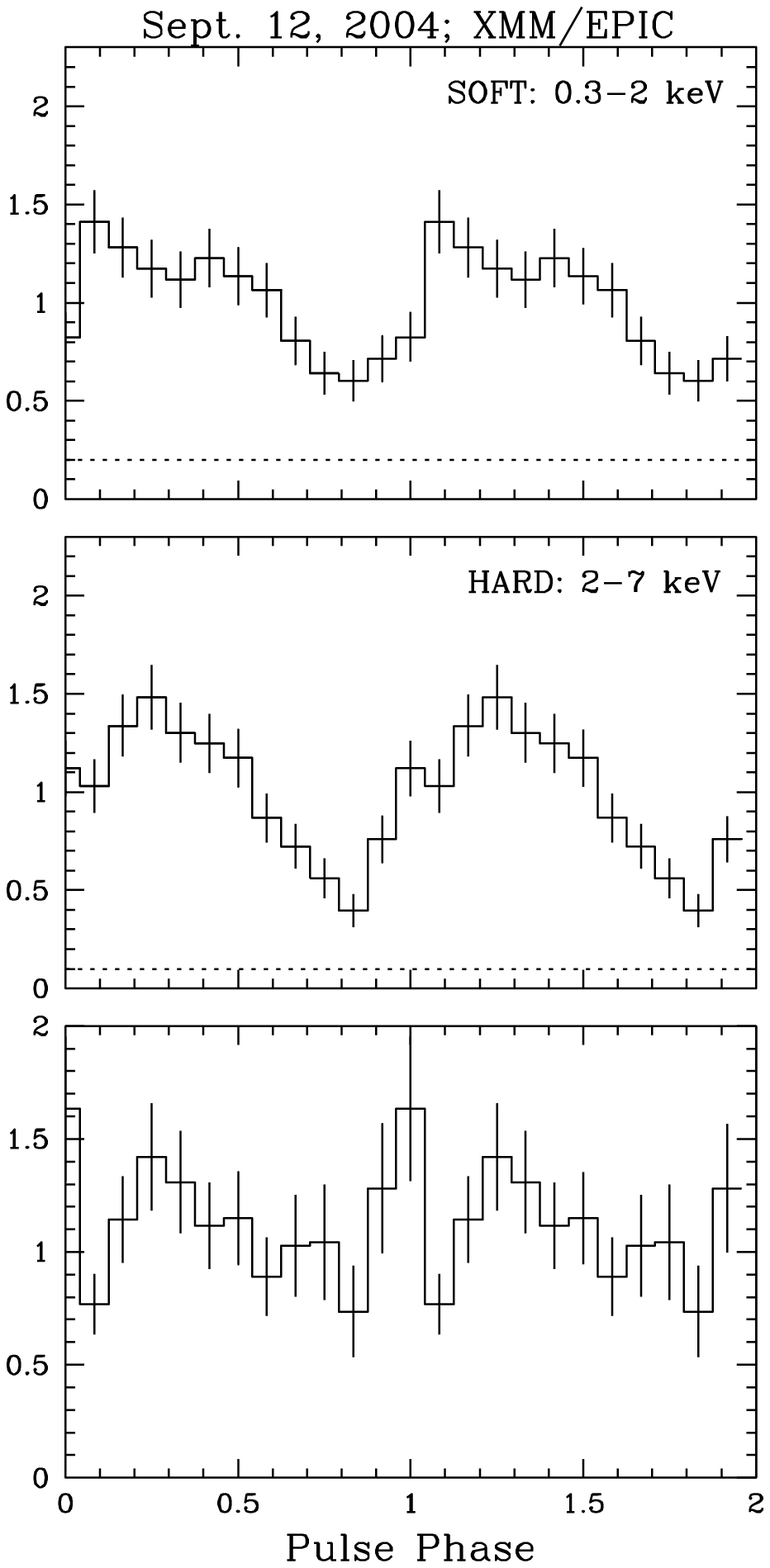}
\caption{X-ray lightcurves of CXOU J073709.1+653544 during the 2004 Aug. 9 and 23 {\em Chandra} and Sept. 12 
{\em XMM-Newton} observations folded at the corresponding best periods (Table \ref{timing_spec_par}) in the soft (0.3-2 
keV) and hard (2-7 keV) energy bands ({\em upper middle panels}) along with hardness ratios, computed taking background 
contribution into account ({\em bottom panels}). The background levels are represented by the dotted lines. 
\label{mod_energy_depend}}
\end{figure}

To investigate the energy dependence and time evolution of the source pulse profile, light curves in the soft (0.3-2 keV) 
and hard (2-7 keV) bands were created for observations \#4, 5, and 6 \footnote{The limited source count statistics 
in the Oct. 3 observation does not permit a detailed study of energy dependence of pulsations.} and folded at the 
corresponding best pulsation periods (Figure \ref{mod_energy_depend}). The corresponding background-corrected pulsed 
fractions in the soft and hard energy bands are given in Table \ref{timing_spec_par}. As can be seen from Fig. 
\ref{mod_energy_depend} and Table \ref{timing_spec_par}, the energy dependence of the pulse profile in the Aug. 9 
{\em Chandra} observation (obs. \#4) is qualitatively different from subsequent {\em Chandra} and {\em XMM-Newton} 
observations (obs. \#5 and 6). The source modulation amplitude in the soft band is higher than in the hard band during 
Aug. 9 {\em Chandra} observation, while the modulation is stronger in the hard band during the Aug. 23 {\em Chandra} and 
Sept. 12 {\em XMM-Newton} observations. The modulation fraction in the soft band is highest (78$\pm$5\%) during the Aug. 9 
observation, and drops to 43$\pm$6\% and 50$\pm$6\% on Aug. 23 and Sept. 12 (Table \ref{timing_spec_par}). At the same time, 
the modulation fraction in the hard energy band remains compatible during all three observations (Table \ref{timing_spec_par}; 
Fig. \ref{mod_energy_depend}).           

\subsection{Long-term Period Evolution}
\begin{figure}
\epsscale{1.0}
\plotone{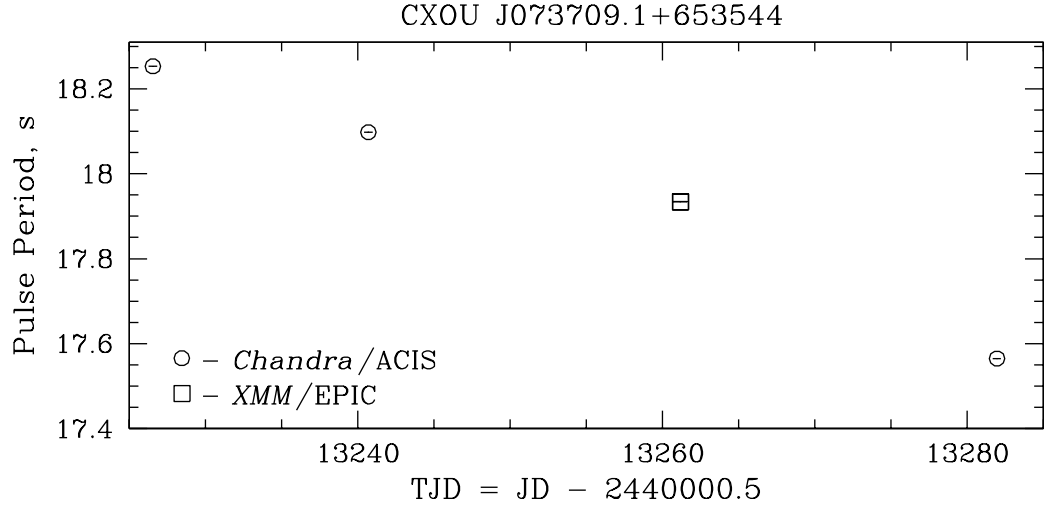}
\caption{The long-term evolution of the pulsation period of CXOU J073709.1+653544. No orbital corrections have 
been applied to the pulse frequencies, since the orbital parameters are unknown. \label{period_evolution}}
\end{figure}

There is a significant change of the pulsation period of CXOU J073709.1+653544 measured over the course of four 
observations covering a $\sim$55-day period from P=18.253 s on Aug. 9, 2004 to P=17.934 s on Sept. 12, 2004 and possibly 
P=17.565 s on Oct. 3, 2004 (Table \ref{timing_spec_par}; Fig. \ref{period_evolution}). During the first three observations 
corresponding to a $35$-day time interval, the source shows a roughly linear decrease of the pulsation period. Using these 
three period measurements and assuming its linear time dependence, we estimate the rate of period change 
$\dot{P}\sim-(1.1\pm0.2)\times10^{-7}$ s s$^{-1}$ (pulse frequency derivative $\dot{\nu}\sim3.4\times10^{-10}$ Hz s$^{-1}$), 
that corresponds to $(\dot{P}/{P})\sim-6\times10^{-9}$ s$^{-1}\sim-0.19$ yr$^{-1}$ for a period $P=18$ s. 

\subsection{X-ray Spectra}
The pulse phase averaged {\em Chandra}/ACIS-S and {\em XMM-Newton}/EPIC spectra of CXOU J073709.1+653544 can be 
adequately fit with the absorbed simple power law models with photon index, $\Gamma \sim 0.9-1.2$ and an equivalent 
hydrogen density N$_{\rm H}\sim(15 - 58)\times10^{20}$ cm$^{-2}$. The corresponding absorbed luminosity of the source 
in the 0.3-7 keV band ranges between $\sim 5\times10^{37}$ and $\sim 2.3\times10^{38}$ ergs s$^{-1}$, assuming the 
distance of 3.2 Mpc. The best-fit spectral model parameters for individual observations are given in Table 
\ref{timing_spec_par}. The luminosity and energy spectra of the source are similar to that of the bright X-ray pulsars 
in our Galaxy and the Magellanic Clouds (Nagase 1989; Yokogawa et al. 2003). The shape of the energy spectra of 
CXOU J073709.1+653544 in the 0.3-7 keV energy band remains essentially constant throught the outburst despite 
significant change of the source luminosity. For all observations, the measured absorbing column $N_{\rm H}$ is 
$\sim$3-10 times higher than the Galactic hydrogen column in the direction of NGC 2403, 4$\times10^{20}$ cm$^{-2}$ 
\cite{DL90}, consistent with an additional intrinsic absorption within the system and inside the disk of NGC 2403. 

\section{DISCUSSION}
A combination of transient behavior and the extremely large spin-up rate of CXOU J073709.1+653544 makes it a unique 
object among pulsating X-ray sources. The overall X-ray properties (X-ray spectrum, pulsation period etc.) of the source 
are consistent with that of the accreting X-ray pulsar in a binary system (White, Swank \& Holt 1983; Nagase 1989). The 
absence of bright optical counterparts (both during quiescence and outburst), transient behavior, overall X-ray properties 
of CXOU J073709.1+653544 and positional coincidence with NGC 2403, allow us to conclude that is almost certainly located 
outside our Galaxy and belongs to NGC 2403. Placing CXOU J073709.1+653544 at the distance of NGC 2403 (3.2 Mpc) implies 
very high luminosities of the source during the outburst ($(0.6\sim2.4)\times10^{38}$ ergs s$^{-1}$ in the 0.3-7 keV energy 
band and $(0.8\sim3.8)\times10^{38}$ ergs s$^{-1}$ in the 0.3-10 keV energy band) with maximum observed luminosity exceeding 
the isotropic Eddington luminosity limit for a $1.4 M_{\odot}$ object accreting hydrogen-rich material. If 
CXOU J073709.1+653544 is indeed an accreting X-ray pulsar in NGC 2403, the relative rate of change of its pulsation period, 
$(\dot{P}/P)\sim-6\times10^{-9}$ s$^{-1}$ is the highest observed in this class of objects to date (Nagase 1989; Bildsten et 
al. 1997). 

If the X-ray pulsations observed in CXOU J073709.1+653544 result from rotation of a highly magnetized accreting compact 
object in a binary system, the change of its pulsation period can be explained by combination of binary orbital motion of the 
pulsar (Skinner et al. 1982) and transfer of angular momentum in the accretion process (Rappaport \& Joss 1977; Ghosh \& Lamb 
1979). The relatively long time scale on which we observe the decrease of the pulsation period and its high, nearly constant rate 
suggest that the intrinsic spin-up of the pulsar should make the major contribution to the observed change of the pulsation period. 
Using the observed pulsation parameters of CXOU J073709.1+653544, and assuming that disk accretion is occurring in the system, 
one can make a simple estimate of the lower limit on the accretion rate required to explain the observed spin-up rate $\dot{P}$. 
Let us assume that all the specific angular momentum of the accreting material is transferred to the compact object at the 
magnetospheric radius $r_{m}$. In order to keep the ``propeller'' effect from preventing accretion onto compact object, the 
magnetospheric radius must be less that the corotation radius $r_{m}\lesssim r_{c}=(GM_{X}/4\pi^{2})^{1/3}P^{2/3}$, where $P$ 
is the rotational period and $M_{X}$ is a compact object mass \cite{IS75}. The maximum torque experienced by compact object 
$N=\dot{M}(GM_{X}r_{c})^{1/2}$, where $\dot{M}$ denotes the accretion rate through the disk, should be greater or equal to the 
observed torque $N_{obs}=I\dot{\omega}=2\pi I(\dot{P}/P^{2})$, where $I$ is the moment of inertia of the compact object. As a 
result, we obtain
\begin{equation}
\dot{M}\gtrsim I\,(4\pi^{2}/GM_{X})^{2/3}\,\dot{P}\,P^{-7/3}
\end{equation}
or,
\begin{equation}
\dot{M}\gtrsim 5.2\times10^{18}\,f\,\left(\frac{-\dot{P}}{10^{-7}\,{\rm s\,s}^{-1}}\right )\left(\frac{P}{18\,{\rm s}}\right )^{-7/3}\; {\rm g\;s}^{-1}, 
\end{equation}
where $f=(I/10^{45}\,{\rm g\,cm}^{2})\,(M_{X}/M_{\odot})^{-2/3}$. Substituting typical white dwarf parameters ($I\sim10^{50}$ g cm$^{2}$, 
$M_{X}=M_{\odot}$) and the observed rotation period and its derivative into equation (2) gives an unreasonably high mass accretion rate 
$\dot{M} \gtrsim 5\times 10^{23}$ g s$^{-1}$. On the other hand, for a typical neutron star parameters ($I\sim10^{45}$ g cm$^{2}$, 
$M_{X}=1.4M_{\odot}$) the estimated mass accretion rate required to produce the observed spin-up is $\dot{M}\gtrsim 4\times 10^{18}$ g 
s$^{-1} = 6.3\times 10^{-8}$ $M_{\odot}$ year$^{-1}$. The corresponding bolometric luminosity of the 1.4$M_{\odot}$ neutron star 
accreting at such rate should be $L_{bol}\gtrsim 7.3\times 10^{38}$ ergs s$^{-1}$, assuming that all gravitational energy of the infalling 
matter released in the process of accretion is converted into radiation. The estimated bolometric luminosity of the source is very high, 
exceeding the Eddington critical luminosity for a 1.4$M_{\odot}$ star accreting hydrogen-rich material at least by factor of 4. 

It is interesting to compare the luminosity required from accretion spin-up mechanism to the total X-ray luminosity of the source 
during the outburst. In order to estimate the total X-ray luminosity of CXOU J073709.1+653544, we extrapolated the results of 
our spectral analysis to the higher energies, assuming that the broad-band spectrum of the source in the 0.3-100 keV band is 
represented by a power law with high-energy cutoff at energies above 10-20 keV, as typically observed in the accreting X-ray pulsars 
(White, Swank \& Holt 1983; Nagase 1989). To model the broad-band spectrum, we used a simple power law model with photon index and 
normalization derived from spectral fitting in the 0.3-7 keV band, modified at energies above a high-energy cutoff $E_{c}$ by the 
function $exp[(E_{c}-E)/E_{f}]$, where $E_{f}$ is the folding energy. For the Aug. 23 observation, the estimated X-ray luminosity 
of CXOU J073709.1+653544 in the 0.3-100 keV energy band falls between $\sim8\times 10^{38}$ and $1.2\times 10^{39}$ ergs s$^{-1}$ 
for $E_{c}=10\sim20$ keV and $E_{f}=10$ keV. Note that this estimate should be regarded as a lower limit, as it does not account for 
the possible presence of the soft ($kT \sim 0.1-0.2$ keV) component in the spectrum of the source, similar to that observed in some of 
the bright X-ray pulsars \cite{Yokogawa03}. Therefore, the estimated maximum X-ray luminosity of CXOU J073709.1+653544 is high 
enough to account for the observed spin-up rate, assuming that the X-ray source is powered by disk accretion onto highly magnetized 
neutron star.

The estimated maximum luminosity of the source can be used to put an upper limit on the strength $B_{0}$ of the magnetic field of the 
neutron star. Assuming the dipole configuration of the pulsar magnetic field, and requiring the magnetospheric radius to be smaller 
than corotation radius $r_{m}\lesssim r_{c}$, we obtain
\begin{equation}
B_{0}\lesssim 7\times10^{13}\,f_{1}\,\left(\frac{P}{18\,{\rm s}}\right )^{7/6}\left(\frac{L_{X}}{10^{39}\,{\rm ergs\,s}^{-1}}\right )^{1/2}\, {\rm G}
\end{equation}
where $f_{1}=(R/10^{6}\,{\rm cm})^{-5/2}\,(M_{X}/M_{\odot})^{1/3}$ and $R$ is the radius of the neutron star.

Although both the observed luminosity of the source and luminosities required by spin-up due to disk accretion are extremely high, they 
lie within the range observed for pulsating neutron stars that can reach luminosities of $\sim 5-10\times 10^{38}$ ergs s$^{-1}$ (White 
\& Carpenter 1978; Skinner et al. 1982; Nagase 1989). The required luminosities are also consistent with theoretical predictions for 
super-Eddington accretion onto highly magnetized ($B \gtrsim 10^{12}$ G) neutron star (Basko \& Sunyaev 1976). At high accretion 
rates, the accreting matter is expected to release its gravitational energy inside accretion columns above the magnetic poles of the 
neutron star. Since most of radiation is emitted from the sides of these columns, it can escape the system without crossing the bulk of 
the inflowing accreting matter. This anisotropy allows the luminosity of the neutron star to exceed formal Eddington limit by large 
factors. For the accretion column with a cross-section in the form of a thin ring segment of length $l$ and thickness $d$, the limiting 
luminosity that can be reached in the Basko \& Sunyaev (1976) model is given by
\begin{equation}
L_{max}\sim 8\times10^{38}\,\left ( \frac{l/d}{40} \right )\,\left ( \frac{M_{X}}{M_{\odot}} \right )\; {\rm ergs \; s}^{-1}
\end{equation}   
Therefore, for the highly super-Eddington accretion rate and $l \gg d$ the emission from accretion columns could explain the 
observed high luminosity of CXOU J073709.1+653544.   

The measured pulse period places CXOU J073709.1+653544 among Be systems or short-period binary supergiant systems with 
Roche lobe overflow on a Corbet diagram (pulse period-orbital period plot)(Corbet 1986; Waters \& van Kerkwijk 1989; Bildsten et al. 
1997). The transient behavior of the source suggests a higher probability for it to be a member of Be binary class, since majority 
of Be systems display recurrent/transient outbursts. The high peak luminosity of CXOU J073709.1+653544 ($L_{X} \sim 10^{39}$ ergs 
s$^{-1}$) is also not unusual for the Be X-ray pulsars with transient Be X-ray pulsar A0538-66 reaching similar luminosities during 
its giant outbursts \cite{Skinner82}. 

Using the Corbet diagram, one can try to predict the value of the binary period of CXOU J073709.1+653544. If it is indeed a Be system, 
the estimated orbital period $P_{orb}$ can vary between $\sim 20$ and $\sim 100$ days. The observed duration of the source outburst 
allows us to put further constraints on the value of the binary period. Assuming that we observed a single outburst from the source and 
that it corresponds to one orbital cycle, we obtain $60 \lesssim P_{orb} \lesssim 100$ days. Note, however, that if the source went through 
the giant (class II) outburst lasting for several orbits (Stella, White \& Rosner 1986) or a sequence of shorter outbursts during 
August-October 2004, the orbital period of the system could be much shorter than 60 days. 

Another possibility is that CXOU J073709.1+653544 is a transient low-mass binary system. The transient behavior and high luminosity 
of the source are somewhat similar to that of the Galactic bursting X-ray pulsar GRO J1744-28 (Kouveliotou et al. 1996), that reached 
a luminosity of $\sim 10^{39}$ ergs s$^{-1}$ during the peak of its outburst (Giles et al. 1996; Sazonov, Sunyaev \& Lund 1997). In the 
low-mass system scenario, the transient outbursts of the source could be explained by sudden increase of accretion from the disk reservoir 
(Sunyaev \& Shakura 1977) or as a result of the viscous-thermal instability of the quiescent accretion disk.    

The discovery of X-ray pulsations in CXOU J073709.1+653544 is yet another demonstration that deep monitoring observations with 
{\em Chandra} and {\em XMM-Newton} have a great potential in revealing populations of both persistent and transient pulsating X-ray 
sources in nearby galaxies. Future observations of CXOU J073709.1+653544, if it reappears, could possibly give deeper insight 
into the behavior of this and other similar X-ray pulsar systems. The complete coverage of the source outbursts with more frequent 
sampling would allow to estimate the orbital parameters of the system and help to determine its nature. 

\acknowledgements
We would like to thank the referee for comments that improved the paper. Support for this work was provided through NASA Grant NAG5-12390. 
This research has made use of data obtained through the High Energy Astrophysics Science Archive Research Center Online Service, provided 
by the NASA/Goddard Space Flight Center. XMM-Newton is an ESA Science Mission with instruments and contributions directly funded by 
ESA Member states and the USA (NASA). The Digitized Sky Surveys were produced at the Space Telescope Science Institute under U.S. Government 
grant NAG W-2166. The Second Palomar Observatory Sky Survey (POSS-II) was made by the California Institute of Technology with funds from the 
National Science Foundation, the National Geographic Society, the Sloan Foundation, the Samuel Oschin Foundation, and the Eastman Kodak 
Corporation.

\clearpage

\begin{table}
\small
\caption{{\em Chandra} and {\em XMM-Newton} Observations of NGC 2403 Used in the Analysis. 
\label{obslog}}
\tiny
\begin{tabular}{ccccccccl}
\hline
\hline
Obs.\# & Date, UT & Date, TJD  & Obs. ID  & Mission/Instrument & Mode/ & RA (J2000)\tablenotemark{a} & Dec (J2000)\tablenotemark{a} & Exp.\tablenotemark{b}\\
       &          &            &          &                    & Filter & (h:m:s)          & (d:m:s)           & (ks)      \\             
\hline
1 &2001 Apr. 17 & 12016.401(0.221) & 2014       & {\em Chandra}/ACIS-S & Faint      &07:36:49.86&+65:34:43.5& 35.5\\
2 &2003 Apr. 30 & 12759.756(0.090) & 0150651101 & {\em XMM-Newton}/EPIC& Full/Thin  &07:36:51.00&+65:36:10.0&  4.7(MOS)\tablenotemark{c}\\
3 &2003 Sep. 11 & 12893.044(0.065) & 0150651201 & {\em XMM-Newton}/EPIC& Full/Thin  &07:36:51.00&+65:36:10.0&  8.9(MOS)/5.0(pn)\\
4 &2004 Aug. 09 & 13226.561(0.273) & 4627       & {\em Chandra}/ACIS-S & Faint      &07:37:22.34&+65:36:10.7& 40.0\\
5 &2004 Aug. 23 & 13240.701(0.290) & 4628       & {\em Chandra}/ACIS-S & Faint      &07:37:21.72&+65:36:10.9& 46.5\\
6 &2004 Sep. 12 & 13261.182(0.449) & 0164560901 & {\em XMM-Newton}/EPIC& Full/Medium&07:37:17.00&+65:35:57.8& 59.0(MOS)/52.4(pn)\\
7 &2004 Oct. 03 & 13281.951(0.276) & 4629       & {\em Chandra}/ACIS-S & Faint      &07:37:19.45&+65:36:29.8& 44.5\\
8 &2004 Dec. 22 & 13361.938(0.304) & 4630       & {\em Chandra}/ACIS-S & Faint      &07:37:12.85&+65:36:20.9& 49.9\\
\hline
\end{tabular}

\tablenotemark{a}{pointing coordinates}\\
\tablenotemark{b}{instrument exposure used in the analysis}\\
\tablenotemark{c}{the source position coincides with the gap between EPIC-pn CCDs, only EPIC-MOS data was used in the analysis}
\end{table}

\clearpage
\begin{table}
\caption{X-ray Pulsation Parameters and Power Law Spectral Fit Information for CXOU J073709.1+653544. 
\label{timing_spec_par}}
\tiny
\begin{tabular}{cccccccccccccl}
\hline
\hline
Obs. & Period &PF$_{0.3-7 keV}$&PF$_{0.3-2 keV}$&PF$_{2-7 keV}$&N$_{\rm H}$&Photon&Flux\tablenotemark{b}& Flux\tablenotemark{c}&$\chi^{2}$&$L_{\rm X}$\tablenotemark{d}&$L_{\rm X}$\tablenotemark{e}&Instrument\\
          &  (s)   & (\%)\tablenotemark{a}     & (\%)\tablenotemark{a}     & (\%)\tablenotemark{a}   &($\times 10^{20}$ cm$^{-2}$)&Index     &          &           &(d.o.f)   &     &    &          \\
\hline
4& $18.25247(73)$ & $70\pm4$ & $78\pm5$& $63\pm6$&$37^{+8}_{-7}$  &$1.18^{+0.13}_{-0.12}$&$1.58\pm0.07$& $1.96\pm0.12$&$25.6(25)$& 1.94 & 2.40 &ACIS-S\\       
5& $18.09825(53)$ & $46\pm4$ & $43\pm6$& $56\pm5$&$20\pm5$        &$0.96\pm0.10$         &$1.90\pm0.07$& $2.12\pm0.14$&$27.7(33)$& 2.33 & 2.60 &ACIS-S\\
6& $17.93443(30)$ & $51\pm3$ & $50\pm6$& $64\pm4$&$15^{+5}_{-4}$  &$0.88\pm0.08$         &$0.89\pm0.04$& $0.99\pm0.06$&$52.4(64)$& 1.09 & 1.21 &EPIC/pn+MOS\\
7& $17.5645(14)$ & ...      & ...     & ...     &$58^{+28}_{-23}$&$1.21^{+0.37}_{-0.34}$&$0.38\pm0.03$& $0.50\pm0.06$&  $3.0(4)$& 0.47 & 0.61 &ACIS-S\\
\hline
\end{tabular}
\tablenotetext{a}{pulsed fraction in the $0.3-7$, $0.3-2$ and $2-7$ keV energy bands, defined as 
(I$_{\rm max}$-I$_{\rm min}$)/(I$_{\rm max}$+I$_{\rm min}$), where I$_{max}$ and I$_{min}$ 
represent source background-corrected intensities at the maximum and minimum of the pulse profile}
\tablenotetext{b}{absorbed model flux in the $0.3 - 7$ keV energy range in units of $10^{-13}$ 
erg s$^{-1}$ cm$^{-2}$}
\tablenotetext{c}{unabsorbed model flux in the $0.3 - 7$ keV energy range in units of $10^{-13}$ 
erg s$^{-1}$ cm$^{-2}$}
\tablenotetext{d}{absorbed luminosity in the $0.3 - 7$ keV energy range in units of $10^{38}$ 
erg s$^{-1}$, assuming the distance of 3.2 Mpc}
\tablenotetext{e}{unabsorbed luminosity in the $0.3 - 7$ keV energy range in units of $10^{38}$ 
erg s$^{-1}$, assuming the distance of 3.2 Mpc }
\end{table}
\clearpage

\clearpage
\thispagestyle{empty}
\setlength{\voffset}{-25mm}
\clearpage
\setlength{\voffset}{0mm}

\begin{figure}
\epsscale{1.0}
\plotone{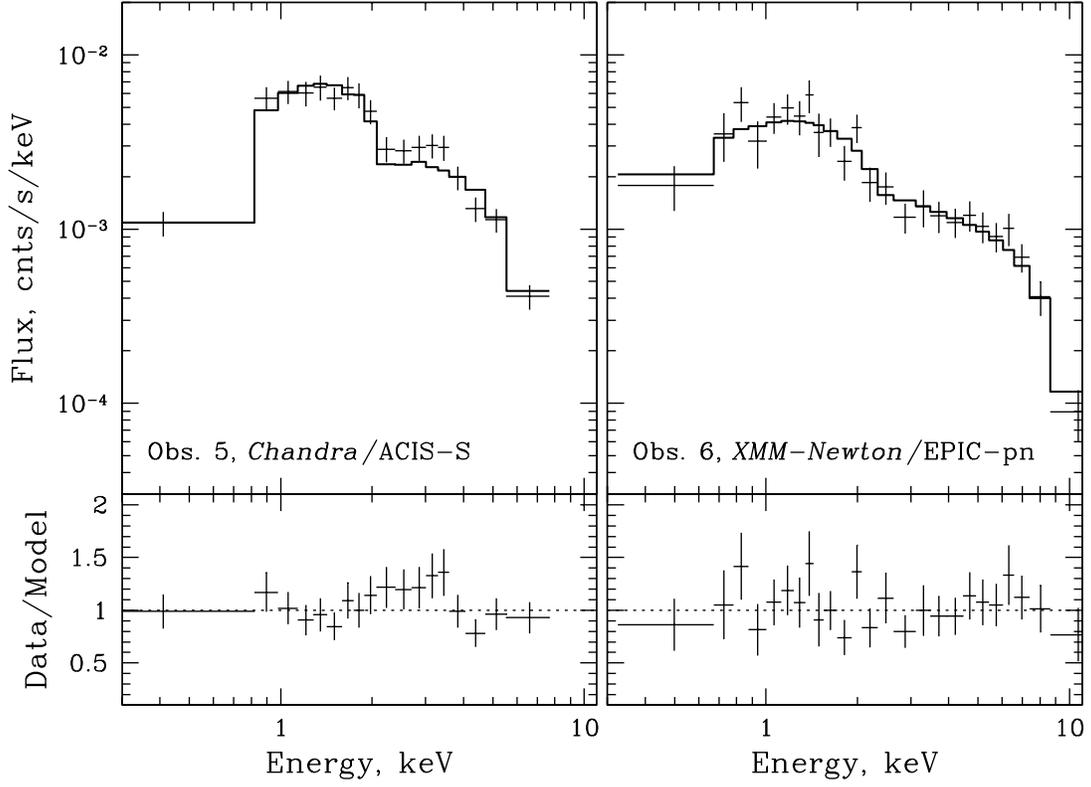}
\caption{{\em Left panels:} Count spectra and model ratios of CXOU J073709.1+653544 obtained during August 23 {\em Chandra}/ACIS-S 
observation. {\em Right panels:} EPIC-pn count spectra and model ratios of the source during September 12 {\em XMM-Newton} observation. 
The best-fit absorbed power law model approximation is shown with thick histograms. \label{spec_fig}}
\end{figure}

\end{document}